 \documentclass[epj]{svjour}

\usepackage{graphics}
\begin{document}

\title{ Spectroscopy of orbital ordering in La$_{0.5}$Sr$_{1.5}$MnO$_4$ : \\ 
A many-body cluster calculation }

\author{Alessandro Mirone\inst{1}, S.S. Dhesi\inst{2}, G. van der Laan\inst{3} }
\institute{ \inst{1}European Synchrotron Radiation Facility, BP 220, F-38043 Grenoble Cedex, France\\
\inst{2} Diamond Light Source,Chilton, Didcot, OX11 0DE, United Kingdom \\
\inst{3}CCLRC Daresbury Laboratory, Warrington WA4 4AD, United Kingdom }

\date{Received: date / Revised version: date}

\abstract{We have studied the orbital ordering (OO) in La$_{0.5}$Sr$_{1.5}$MnO$_4$ and its
soft x-ray resonant diffraction spectroscopic signature at the Mn $L_{2,3}$ edges. We have
modelled the system in second quantization as a small planar cluster consisting of a central
Mn atom, with the first neighbouring shells of oxygen and Mn atoms. For the effective
Hamiltonian we consider Slater-Koster parameters, charge transfer and electron correlation
energies obtained from previous measurements on manganites. We calculate the OO as a function
of oxygen distortion and spin correlation used as adjustable parameters. Their contribution
as a function of temperature is clearly distinguished with a good spectroscopic agreement.
\PACS{
{61.10.-i}{     X-ray diffraction and scattering} --
{71.30.+h}{     Metal-insulator transitions and other electronic transitions} --
{71.10.-w}{     Theories and models of many-electron systems} --
{78.20.Bh}{     Theory, models, and numerical simulation}
}
}
\authorrunning{A. Mirone et al.}
\titlerunning{Spectroscopy of orbital ordering in La$_{0.5}$Sr$_{1.5}$MnO$_4$}

\maketitle

\section{Introduction}

Recent soft x-ray resonant diffraction measurements on half-doped manganite
La$_{0.5}$Sr$_{1.5}$MnO$_4$ have been interpreted in terms of orbital
ordering (OO) \cite{Dhesi,Wilkins,Staub}. There is a strong interest in understanding how OO
settles in this system as a result of the interplay between charge, orbital and spin degrees
of freedom. In fact, this interplay can be found in many other different phenomena, such as
high-temperature superconductivity, colossal magnetoresistance and magnetostructural
transitions
\cite{osborne}.

The MnO$_2$ plane with its superstructure for half doping at low temperature is shown in
Figure 1 \cite{Radaelli,Mutou,Khomskii,Hotta,Daoud,Stojic}. The two charge-separated Mn sites,
which we shall denote by Mn$^{3+}$ and Mn$^{4+}$ (although the charge separation is
fractional) display a checker board alternation. One can see ferromagnetic zig-zag chains,
where the Mn$^{4+}$ sites form the corners and the Mn$^{3+}$ sites are in the middle of the
straight segments. Adjacent zig-zag chains are antiferromagnetically aligned with respect to
each other. There is a distortion of the oxygen atoms consisting of an elongation of the
Mn$^{3+}$-O bonds along the zig-zag segments. Figure 1 shows the occupied Mn$^{3+}$ $e_g$
orbital under the hypothesis of $3z^2$$-$$r^2$/ $3x^2$$-$$r^2$ ordering, which is a possible
simplified way of looking at the electronic structure, although so far there is not yet a
consensus regarding the correct ordering. More importantly, there is no consensus yet
regarding the way the OO and spin-ordering settle in the system as a function of temperature.

Based on recent soft x-ray resonant diffraction experiments we will try to answer these still
open questions with a theoretical analysis of the spectra.
The soft x-ray resonant diffraction experiments that we consider have been done on
La$_{0.5}$Sr$_{1.5}$MnO$_4$ at the Mn $L_{2,3}$ edges where the scattering factors exhibit
huge variations with a strong dependence on the precise $3d$ orbital occupation and ordering.

Considering the stoichiometry in La$_{0.5}$Sr$_{1.5}$MnO$_4$, the average occupation of the
Mn $3d$ shell can be estimated at about 3.5 electrons. The soft x-ray diffraction spectra
show a strong multiplet structure due to transitions $3d^n \leftrightarrow
2p^53d^{n+1}$~\cite{laan} which are complicating the analysis. Previous studies
\cite{Dhesi,Wilkins,Staub} simplified the analysis of the multiplet structure by considering
integer charge ordering (CO) with a
$3d$ occupation of 4 electrons on the active sites (active in the OO diffraction process). The
scattering factors of the active site (Mn$^{3+}$) were then calculated in the crystal-field
approximation. Although such calculations were able to produce some reminiscence of the
observed spectra, the agreement was not good enough to clearly discriminate between the
different hypotheses concerning the occupied orbitals. Moreover, these calculations were
based on a local mean-field approximation and were inherently unable to treat the effects of
spin correlation between the different Mn sites. The approximation of integer occupation is
contrasting with band calculations results, which estimate the charge separation between
Mn sites to be about 0.3 (Ref.~\cite{mizokawa}) instead of one electron.

\begin{figure} 
\rotatebox{0}{
\resizebox{0.75\columnwidth}{!}{
\includegraphics{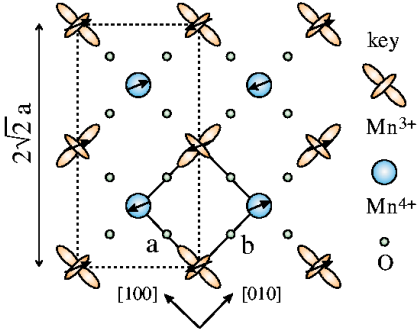}
}
}
\caption{The MnO$_2$ plane and its super structure in the case of half doping at low
temperature.}
\end{figure}

In this work we will go beyond the limits of previous analyses, by building a cluster around
the active Mn site. We consider a model where the degenerate $3d$ shell of the active site is
coupled to the neighbouring oxygen orbitals by hopping terms. On their
turn, the oxygen orbitals are coupled to the $3d$ orbitals of the inactive Mn$^{4+}$ sites
(these for symmetry reasons inactive sites do not contribute to the OO diffraction). For the
inactive Mn sites we consider only one $e_g$ orbital. For this orbital we add a
spin-dependent energy term to simulate the spin magnetisation. In this model we can play with
two main ingredients, i.e., the distortion of the oxygen positions (Jahn-Teller distortion)
which rescales the hopping, and the spin magnetisation of the inactive sites.

In Sec.~\ref{sec:Hamiltonian} we define the Hamiltonian and in Sec.~\ref{sec:Parameter} we
fix the parameters. In Sec.~\ref{sec:Results} we compare the experimental spectra to the
calculated spectra obtained by choosing an optimal distortion. The study of the dependence of
the calculated spectra on distortion and spin magnetisation is summarized in
Sec.~\ref{sec:Conclusions}.

\section{Model Hamiltonian}
\label{sec:Hamiltonian}
We consider a model where the degenerate $3d$ electrons of the active site are coupled to the
neighbouring oxygen orbitals by a hopping term modulated by Slater-Koster parameters. The
oxygen orbitals are on their turn coupled to the $3d$ orbitals of the inactive Mn$^{4+}$
sites. The Hamiltonian of the total system is
\begin{equation}
        H= H_a + T_1 +T_2 +H_1 + H_2 \, ,
\end{equation}
\noindent
where $H_a$ is the atomic Hamiltonian for the active Mn site and $H_1$ and $H_2$ are the
Hamiltonians for the first-neighbour oxygen atoms and the neighbouring inactive Mn$^{4+}$
sites, respectively, in the absence of hopping. The $T_1$ and $T_2$ are the hopping terms for
the Mn$^{3+}$-O and O-Mn$^{4+}$ bonds, respectively. 
\begin{equation}
T_1 = \sqrt 2 \,t \sum_{\sigma}\left(
          g  \,  o_{x,\sigma}^{\dagger} d_{x^2,\sigma}  +
   o_{z,\sigma}^{\dagger} d_{z^2,\sigma} +
    o_{y,\sigma}^{\dagger} d_{y^2,\sigma}\right) + \mathrm{c.c.},
\end{equation}

\noindent
where the $z$ and $x$ axes lie in the MnO$_2$ plane, $t$ is the Slater-Koster $V_\sigma$
parameter, $g$ is the reduction factor of the hopping along $x$ (which is taken parallel to
the zig-zag segment). 
The $d$ and $o$ are the second-quantization operators for Mn $3d$ and O $2p$ electrons,
respectively. Neglecting the smaller $V_\pi$ parameter, we consider only three oxygen
orbitals (six including spin) and
$o_{x,y,z}$ represents the orbitals in the $x$,$y$,$z$ directions. Only one oxygen per
direction is considered and a factor $\sqrt 2$ is included in $T_1$, so that the $o$'s
represent symmetrized orbitals. The $d_{x^2}$, $d_{y^2}$ and
$d_{z^2}$ are linear combinations of $e_g$ operators pointing along the three cartesian
directions (e.g., $d_{x^2} =
\sqrt 3/2 \, d_{x^2-y^2}+1/2 \, d_{z^2}$). The $o_z$ and $o_x$ degrees of freedom are on their turn
coupled to two Mn$^{4+}$ sites by a hopping term
\begin{equation}
T_2 = t  \sum_{\sigma }\left(
           o_{x,\sigma}^{\dagger} X_{\sigma}  +
         o_{z,\sigma}^{\dagger} Z_{\sigma}  \right) + \mathrm{c.c.},
\end{equation}

\noindent
where $X$ ($Z$) represents an $e_g$ orbital at the Mn$^{4+}$ site along the $X$ ($Z$)
direction.

The Hamiltonian $H_1$ for the isolated oxygen atoms is
\begin{equation}
        H_1 =  \sum_i  [ \epsilon_p  \sum_{\sigma}  o^{\dagger}_{i,\sigma} o_{i,\sigma}  
+ U_{pp} (1- o^{\dagger}_{i\uparrow} o_{i\uparrow})(1- o^{\dagger}_{i\downarrow} o_{i\downarrow})],
\end{equation}

\noindent
with $i \in \{x,y,z\}$. The Hamiltonian $H_2$ is 
\begin{equation}
\label{externalMn}
        H_2  =  \sum_{\sigma} [\epsilon_d   +h (\frac{1}{2} + \sigma_{\acute{z}}) ]
X^{\dagger}_{\sigma} X_{\sigma}  
+ \sum_{\sigma} [\epsilon_d + h (\frac{1}{2} - \sigma_{\acute{z}})  ]
Z^{\dagger}_{\sigma} Z_{\sigma}     \label{h2}, 
\end{equation}

\noindent
where $h$ is the exchange energy term that takes into account the opposite magnetisation of
the two Mn$^{4+}$ sites. The Hubbard correlation term is absent in $H_2$, but in the
calculation we limit the Hilbert space by disregarding states with doubly occupied $X$ or $Z$
orbitals.

\section{Choice of parameters}
\label{sec:Parameter}

The parameters for the atomic Hamiltonian $H_a$, which are the spin-orbit interactions and
Slater integrals -- except for the monopole terms which are strongly screened -- have been
obtained using Cowan's Hartree-Fock code~\cite{Cowan} averaging the values for the Mn$^{3+}$
and Mn$^{4+}$ configurations. The atomic parameters are listed in Table 1. We apply rescaling
factors of $0.75$ and $0.8$ to the $dd$ and $pd$ Slater integrals, respectively, as is common
practise to obtain effective Slater integrals for correlated transition metal
systems~\cite{laan}.  For $F_{dd}^0$ we start from the estimated $U_{dd}$ of about 5 eV from
previous studies~\cite{mizokawa} based on spectroscopical data. Considering a starting
configuration $t^3_{2g\downarrow}e^1_{g\downarrow}$ and adding two extra electrons we obtain
$t^3_{2g\downarrow}e^2_{g\downarrow}t^1_{2g\uparrow}$. Therefore the experimental $U_{dd}$
is the interaction between a $t_{2g\uparrow}$ and an $e_{g\downarrow}$ electron,
\begin{equation}
U_{dd}=F_{dd}^0 - 4/49 \,F_{dd}^2 - 2/147 \,F_{dd}^4,
\end{equation}

\noindent
from which $F_{dd}^0$ is deduced.

The hopping $t$ is estimated as 1.8 eV and $U_{pp}$ as 5 eV. The bare energy $\epsilon_p$ of
the oxygen orbitals is deduced from the charge-transfer energy $\Delta$ = 4 eV. This
experimental parameter is defined as the energy to transfer an oxygen electron onto a bare Mn
atom (in the absence of hybridization). Therefore, 
\begin{eqnarray}
  \epsilon_{p}&= -\Delta + n_d F_{dd}^0  + 6 F_{pd}^0  - 14/49 \,F_{dd}^2 - 14/49 \,F_{dd}^4  \nonumber \\
&- 2/5 \,G_{pd}^1 - 9/35 \,G_{pd}^3  + (2-n_p) U_{pp} + \delta_{\mathrm{hyb}}, 
\end{eqnarray} 

\noindent
where $n_d$ and $n_p$ are the average occupation numbers of the Mn $3d$ and O $2p$ shell,
respectively, in the ground state of the model Hamiltonian and $\delta_{\mathrm{hyb}}$ is
the residual energy shift of the oxygen orbitals when we consider hybridization with inactive
sites only. The value that we find is $\epsilon_p$ = 54.9 eV. Of course, such a large energy
value is not referenced to the vacuum level but to a bare $3d$ orbital with zero occupation.
The spin-magnetisation parameter $h$ for the Mn$^{4+}$ sites is 2.5 eV in the case of full
spin ordering. The energy of the bare orbitals $X$ and $Z$ is
\begin{equation}
  \epsilon_d = \epsilon_p + \Delta + \delta_d \, ,
 \end{equation}  

\noindent
where $\delta_d$ is a free parameter that is allowed to take values in the order of 1 eV,
which accounts for different effects, such as the charge splitting between the sites and an
additional effective ligand field that compensates for the incomplete inclusion of
hybridization for the $X$ and $Z$ orbitals in the model. The spin quantization axis
$\acute{z}$ in Eq.~\ref{externalMn},  which fixes the orientation of the magnetisation for
the Mn$^{4+}$ sites, lies in the (001) plane and we have tested several different
orientations.

\begin{table*}
\begin{tabular}{lcccccccccc}
\hline
Configuration     &&   $F_{dd}^2$ &  
$F_{dd}^4$ & $\zeta_{3d}$ & $F_{pd}^2$ & $G_{pd}^1$ & $G_{pd}^3$ &  $\zeta_{2p}$ &     
$F_{dd}^0$ & $F_{pd}^0$     \\
\hline 
Ground state  && $8.94$ & $5.62$ & $0.051$ & $5.86$    & $4.38$    & $2.5$   & $-$  & 
$5.7$  &  $1.1 F_{dd}^0$  \\
Excited state && $9.53$ & $5.98$ & $0.063$ & $5.86$ & $4.38$ & $2.5$ & $6.85$ &    
$5.7$  & $1.1 F_{dd}^0$ \\
\hline
\end{tabular}
\vspace{10pt}
\caption{\label{atomicpar} Calculated atomic Hartree-Fock values~\cite{Cowan} for the
configuration averaged Slater integrals and spin-orbit interactions (in eV).}
\end{table*}

\section{Results}
\label{sec:Results}

\begin{figure} 
\rotatebox{-90}{
\resizebox{0.8\columnwidth}{!}{
\includegraphics{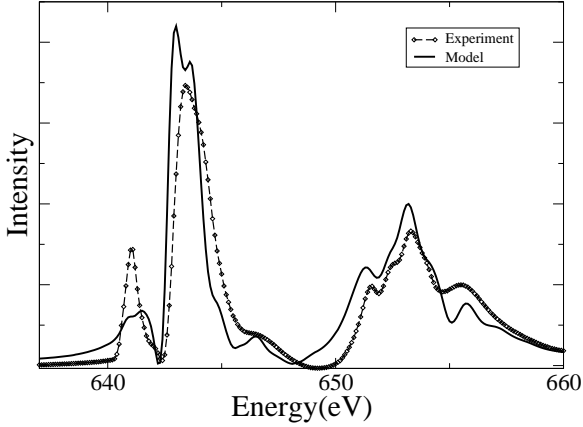}
}
}
\caption{Experimental diffraction spectrum (dashes with diamonds) and calculation (solid line)
for an optimal distortion and correlation.}
\end{figure}

\begin{figure} 
\rotatebox{-90}{
\resizebox{0.8\columnwidth}{!}{
\includegraphics{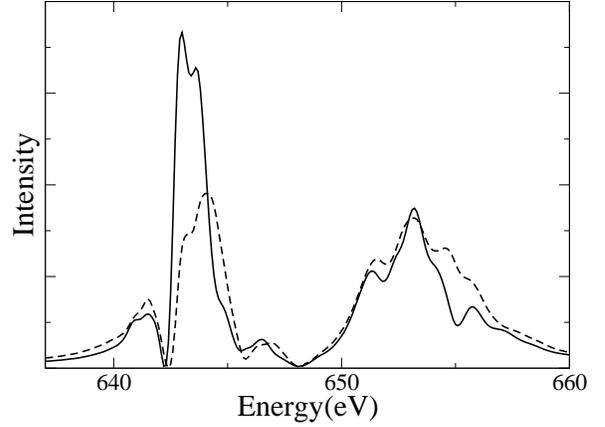}
}
}
\caption{Calculated diffraction spectra for different values of the distortion and magnetic
correlation parameter. Solid line is for $g=0.7$; $h=2.5$. Dashed line is for $g=0.85$;
$h=1.25$.}
\end{figure}

Figure 2 shows the calculated spectrum for a hybridization reduction factor $g=0.7$ and a
Mn$^{4+}$ energy term $\delta_d$ = 0.8 eV. In the calculation the Hilbert space of
the ground-state configuration and the excited-state configuration (with a $2p$ hole and an
extra electron in the valence band) have been fully expanded with the constraint to have
between 8 and 12 electrons on the Mn$^{4+}$ sites and zero or one electron in each of the
$X$, $Z$ orbitals. The scattering factors are calculated in the dipole approximation and using
a Lorentzian line width broadening of 0.45 (0.55) eV for the $L_3$ ($L_2$) edge. The
magnetisation  is taken along the [110] direction which gives the best fit. The reflectivity
is calculated as the squared OO scattering factor divided by the absorption of the
sample~\cite{Dhesi}. The absorption is calculated as a stoichiometric average of the
La$_{0.5}$Sr$_{1.5}$MnO$_4$ elemental absorption. The absorption for Mn is obtained using the
experimental absorption measured for La$_{0.5}$Sr$_{1.5}$MnO$_4$ \cite{Dhesi} at the
$L_{2,3}$ absorption edges and joining it with the tabulated values. The experimental
spectrum has been measured at 134 K
\cite{Dhesi}.

Our calculation reproduces all the spectral features, the only problem is that the
right-hand shoulder of the lowest energy peak has too much intensity and appears as a
separate peak. It is interesting to look at the behavior of the system as a function of the
oxygen displacement. Going from zero distortion ($g\simeq 1$) to strong distortion ($g<0.5$)
we have observed that the $L_3$ peak is already dominant at very low distortion ($g\simeq
1$), while at lower $g$ values it loses intensity compared to the $L_2$ peak. This result
is very different from all previous analyses based on simple ligand-field models~\cite{castelton}, which
predict a direct relation between the $L_3$ main peak intensity and the Jahn-Teller
distortion. We think that the ligand-field model may  fail in this case because
 the resonant diffraction is sensitive to the scattering factor differences for different
polarisations and these differences depend not only on one-particle energy shifts but also
( thinking in terms of one-particle Green functions )
 on the spectral weight transfer due to hybridisation, which
ligand-field model neglects. We think that, as a consequence, ligand-field model calculations
tend to give stronger splitting of $3d$ orbitals  to compensate for the missing spectral weight effect,
in particular  Stojic et. al \cite{Stojic}  find a $1.35$eV splitting for the
$t_{2g}$ band. Such a value seems unusual to us.

We obtained the optimal fit for $g=0.7$. For $g$ values between 1 and 0.6 the magnetisation of
the Mn$^{3+}$ site is parallel to that of the Mn$^{4+}$ site which lies along the $X$
direction. For $g$ values below 0.6 the magnetisation is reversed. What happens is that the
$e_g$ orbital, laying along the $3x^2$$-$$r^2$, aligns its spin parallel to that of the
Mn$^{3+}$ atom which is found in the $Z$  direction because for too small $g$ the effective
hopping on the $X$ orbital becomes weaker than on the $Z$ orbital. However this phenomenon
could be an artefact of our model. In fact, the model neglects the $t_{2g}$ hybridization
which always gives an antiferromagnetic coupling with the $Z$ orbital. In any case beyond
this, possibly artificial, magnetisation-reversal threshold the $L_3$ main peak is
strongly enhanced compared to the $L_2$ and it shifts to lower energy by about 1.5 eV. This
behavior resembles what happens in the case of the Pr$_{0.6}$Ca$_{0.4}$MnO$_3$ system below
$T_{\mathrm{N}}$ \cite{Sawa}. It is interesting to test with the model to see what happens
when the spin correlation is reduced. Our model, in particular the $H_2$ term in
Eq.~\ref{h2}, considers a well established spin order and is simplified by using symmetrized
orbitals. A way to mimic the reduction of the spin correlation, without losing the simplicity
of the model, is reducing the value of $h$ in Eq.~\ref{h2}. We show in Figure 3, with the
solid line, the diffraction spectra for reduced values of the distortion ($g=0.85$) and spin
correlation ($h$ = 1.25 eV). The solid line is the optimal fit to the experimental data (from
Figure 2). The change in peak heights reproduces the experimental behavior between
$T_{\mathrm{C}}$ and $T_{\mathrm{N}}$ \cite{Dhesi}, i.e., the main $L_3$ peak loses intensity
and the right and left shoulders of the $L_2$ gain in intensity.

Analysing the one-particle Green function for the ground state we find that in our model the
$e_g$ electron  spends only $15$\% of its time on the Mn$^{4+}$ site. This is a very
pronounced charge separation. We also find that the occupied $e_g$ orbital at the Mn$^{3+}$
site has $3 x^2$$-$$r^2$ symmetry. This result  contradicts the interpretation of the
experimental x-ray linear dichroism at the Mn $L_{2,3}$ edges by Huang {\it et
al.}~\cite{huang}. They have observed a stronger absorption for in-plane linear polarisation
than for out-of-plane polarisation, from which they deduce that the occupied $e_g$ orbital
at the active sites has $x^2-y^2$ symmetry.

\section{Conclusion}
\label{sec:Conclusions}
We have analysed the experimental soft x-ray resonant diffraction of the orbital ordering in
the half-doped manganite La$_{0.5}$Sr$_{1.5}$MnO$_4$  using a many-body cluster calculation.
The dependence on the order parameters of the peak intensities calculated by our model is
very different -- if not completely opposite -- to those found using simpler ligand-field
models. In our model a central Mn$^{3+}$ site hybridizes with the first shell of neighbouring
oxygen atoms. On their turn the oxygen atoms are hybridized with neighbouring Mn$^{4+}$ atoms
in the MnO$_2$ plane. The experimental spectrum at 134 K has been reproduced with good
agreement, using a slight distortion of the planar oxygens and by supposing a strong local
magnetic correlation between Mn sites. The temperature dependence of the spectra between
$T_{\mathrm{C}}$ and $T_{\mathrm{N}}$ has been reproduced by reducing simultaneously the
oxygen distortion and the spin correlation. It is not possible to reproduce the temperature
dependence of the spectra between $T_{\mathrm{C}}$ and
$T_{\mathrm{N}}$ by simply reducing the distortion. This shows that spin correlation is an
important ingredient between $T_{\mathrm{C}}$ and $T_{\mathrm{N}}$ and increases with the
lattice distortion. We also found a pronounced charge separation between the two sites
and an in-plane orbital polarisation. The latter does not yet allow us to explain the
the results of some recent x-ray linear dichroism observations.


We thank the ESRF computing service for computation of the spectra, in particular 
Wolf Dieter Klotz and Gaby Forstner for taking care of computer clusters and grid computing.

\end{document}